\documentclass[prd,aps,amsmath,twocolumn,superscriptaddress,amssymb,reprint,nofootinbib]{revtex4-2}

\usepackage{graphicx,array}
\usepackage{hyperref}
\usepackage{color}
\usepackage{amsmath,amssymb,slashed,latexsym}
\usepackage{bm}
\usepackage{braket}
\usepackage{placeins}
\usepackage{subcaption}

\definecolor{darkblue}{cmyk}{1,0.4,0,0.3}
\definecolor{violet}{cmyk}{0,1,0,0.2}
\hypersetup{colorlinks, bookmarksnumbered, citecolor=darkblue, linkcolor=darkblue, pdfstartview=FitH, urlcolor=darkblue, linktocpage}

\newcommand{\TeV}{\text{TeV}}
\newcommand{\Br}{\mathcal{B}}

\begin{document}
\title{A Minimal Explanation of Flavour Anomalies:\\ B-Meson Decays, Muon Magnetic Moment, and the Cabibbo Angle}

\author{\large David Marzocca}\email{david.marzocca@ts.infn.it}
\author{\large Sokratis Trifinopoulos}\email{sokratis.trifinopoulos@ts.infn.it}
\affiliation{INFN, Sezione di Trieste, SISSA, Via Bonomea 265, 34136, Trieste, Italy}

\begin{abstract}
Significant deviations from the Standard Model are observed in semileptonic charged and neutral-current B-decays, the muon magnetic moment, and the extraction of the Cabibbo angle. We propose that these deviations point towards a coherent pattern of New Physics effects induced by two scalar mediators, a leptoquark $S_1$ and a charged singlet $\phi^+$. While $S_1$ can provide solutions to charged-current $B$-decays and the muon magnetic moment, and $\phi^+$ can accommodate the Cabibbo-angle anomaly independently, their one-loop level synergy can also address neutral-current $B$-decays.
This framework provides the most minimal explanation to the above-mentioned anomalies, while being consistent with all other phenomenological constraints.
\end{abstract}

\maketitle

\section{Introduction}

The Standard Model (SM) of particle physics provides an exquisite description of the interactions between fundamental particles in a very broad spectrum of energies. There are, however, experimental and theoretical reasons to expect departures from the SM due to some New Physics (NP) sector.
Intriguingly, since a few years certain low-energy flavour measurements pursued at the LHC and several other experiments started exhibiting a number of deviations from SM predictions, that have been growing in significance with the addition of more data. \par
Firstly, there are hints of Lepton Flavor Universality (LFU) violation in semi-leptonic $B$-meson decays: 
\begin{itemize}
	\item \boldsymbol{$b \to c \tau \nu.$} An enhancement of the charged-current transition in $\tau$ vs. light leptons~\cite{Lees:2013uzd, Hirose:2016wfn, Aaij:2015yra, Aaij:2017deq, Abdesselam:2019dgh} with respect to the SM prediction~\cite{Bernlochner:2017jka,Bigi:2017jbd,Jaiswal:2017rve}, as encoded by the ratios
\begin{equation}
R_{D^{(*)}} = \frac{ \mathcal B(B \to D^{(*)} \tau \overline{\nu})}{ \mathcal B(B \to D^{(*)} \ell \overline{\nu} ) }~,
\end{equation}
is observed at approximately $3\sigma$.
	\item \boldsymbol{$b \to s \ell \ell.$} A deficit of the neutral-current transition in muons vs. electrons~\cite{Aaij:2017vbb, Aaij:2019wad, Abdesselam:2019wac, Abdesselam:2019lab, Aaij:2021vac} manifests in the ratios
\begin{equation}
R_{K^{(*)}} = \frac{ \mathcal B(B \to K^{(*)} \mu\overline{\mu}) }{ \mathcal B(B \to K^{(*)} e \bar{e} )}~,
\end{equation}
that is predicted to be equal to 1 with high accuracy in the SM ~\cite{Bordone:2016gaq}.
Remarkably, the update on $R_K$ presented recently by the LHCb collaboration~\cite{Aaij:2021vac}, confirmed the trend observed before and increased the significance of the deviation.
Including also an observed deviation in $\Br(B^0_s \to \mu^+ \mu^-)$ \cite{Aaij:2015esa,Aaij:2017vad,Aaboud:2018mst,Sirunyan:2019xdu}, that can be precisely predicted in the SM, the combined significance of the deviation reaches $4.7\sigma$ \cite{Geng:2021nhg,Altmannshofer:2021qrr,Alguero:2019ptt}.\par
Furthermore, taking into account also less theoretically-clean observables, e.g. differential angular distributions of the $B \to K^* \mu^+ \mu^-$ decay, as well as several branching ratios of $b \to s \mu \mu$ processes~\cite{Aaij:2013qta,Aaij:2015oid,Aaij:2020nrf}, the overall significance of the deviations in this channel is raised to even above $6 \sigma$, depending on the specific SM prediction employed \cite{Ciuchini:2019usw,Alguero:2019ptt,Altmannshofer:2021qrr,Geng:2021nhg}.
\end{itemize}
The other two precision measurements featuring anomalous results are:
\begin{itemize}
	\item \boldsymbol{$(g-2)_{\mu}.$}
	The longstanding deviation  from the SM prediction in the anomalous magnetic moment of the muon $a_\mu = (g-2)_{\mu}/2$ recorded by the BNL experiment~\cite{Bennett:2006fi} has recently been updated by FNAL~\cite{Abi:2021gix}, confirming the previous trend and increasing the significance of the deviation from the SM prediction \cite{Aoyama:2020ynm} to an overall $4.2\sigma$ level.\footnote{See however \cite{Borsanyi:2020mff}, that claims a much reduced discrepancy between SM and measurement, as well as the corresponding discussion in \cite{Aoyama:2020ynm}.}
	\item \textbf{Cabibbo-Angle Anomaly (CAA).} Discrepancies between the different determinations of the Cabibbo angle were reported recently. In particular, the values of  $V_{us}$ extracted from $K \to \pi \ell \nu$ decays, the ratio $\mathcal B(K \to \mu \nu)/\mathcal B(\pi \to \mu\nu)$ and CKM unitarity using the value of $V_{ud}$ estimated by superallowed nuclear $\beta$ decays. The tension amounts to $3.6\sigma$ or $5.1\sigma$~\cite{Belfatto:2019swo,Grossman:2019bzp} depending on the input from the nuclear $\beta$ decays (i.e. Ref. \cite{Czarnecki:2019mwq}  or Ref. \cite{Seng:2018qru}).
\end{itemize}

In this letter, we present the minimal ultraviolet (UV) complete NP framework that can provide a combined explanation to the above-mentioned anomalies while being consistent with all other phenomenological constraints. The relevant particle content consists of the $S_1$ scalar leptoquark (LQ) and the singly charged scalar $\phi^+$,  with quantum numbers under $(\text{SU}(3)_c , \text{SU}(2)_L)_{\text{U}(1)_Y}$:
\begin{equation}
   S_1 \sim (\boldsymbol{\bar{3}},\textbf{1})_{1/3}~, \qquad
   \phi^+ \sim (\textbf{1},\textbf{1})_{1}~.
\end{equation}
The $S_1$ LQ has been considered as a mediator for a simultaneous explanation of $R_{D^{(*)}}$, at tree-level, and $(g-2)_{\mu}$, at one-loop \cite{Bauer:2015knc,Cai:2017wry,Crivellin:2019dwb,Saad:2020ihm,Gherardi:2020qhc,Lee:2021jdr}.
Additionally, the scalar $\phi^+$ modifies the tree-level decay of a charged lepton into a lighter one and a neutrino pair, which in turns translates into a shift of $V_{ud}$ necessary to explain the CAA \cite{Crivellin:2020oup,Crivellin:2020klg,Felkl:2021qdn,Crivellin:2021njn}.
While $S_1$ alone cannot explain completely the neutral-current anomalies $b \to s \ell \ell$ via its one-loop contributions \cite{Cai:2017wry,Azatov:2018kzb,Angelescu:2018tyl,Gherardi:2020qhc,Angelescu:2021lln}, we show that the inclusion of an additional box diagram involving both $S_1$ and $\phi^+$ can achieve a very good fit of the data.
To this end, we stress that the inclusion of the purely leptonic interactions of $\phi^+$, that complement the LQ ones in the full resolution of the B-physics anomalies, is fully compatible with the hints towards LFU violation in $\tau$ decays. 

We notice that the present model is the most economical. This is due to the fact that none of the proposed one- (or two-particle) solutions can address more than two (or three) out of the four flavour anomalies simultaneously.
For instance, the vector LQ models~\cite{Barbieri:2015yvd,Buttazzo:2017ixm,DiLuzio:2017vat,Bordone:2017bld,Calibbi:2017qbu,Blanke:2018sro,Angelescu:2018tyl,Cornella:2021sby} cannot account neither for $(g-2)_{\mu}$ nor CAA, and at least two new particles would be necessary in order to improve the combined fit, while the scalar LQ singlet plus triplet solution~\cite{Crivellin:2019dwb,Saad:2020ihm,Gherardi:2020qhc,Lee:2021jdr} can explain three out of four anomalies without addressing the purely leptonic CAA.

In the following we present the model and perform a global analysis of the anomalous observables and all the relevant constraints, evaluating the improvement over the SM. Finally, we briefly discuss the implications for future experiments.

\section{Model}
\label{sec:model}

The SM Lagrangian is augmented by the following Yukawa-type terms\footnote{In principle, there exist also quartic couplings between the scalars themselves and between the scalars and the Higgs. They are not relevant for the phenomenological analysis of this work and are thus omitted.}
\begin{equation}
\label{eq:S1+phi_Lgr}
\mathcal L_{S1+\phi} = \frac{1}{2} \lambda_{\alpha \beta} \bar{\ell}_{\alpha}^c \epsilon \ell_{\beta} \phi^+ +\lambda_{i \alpha}^{1L} \bar{q}_i^c \epsilon \ell_{\alpha} S_1 + \lambda_{i \alpha}^{1R} \bar{u}_i^c e_{\alpha} S_1 + \rm{h.c.}~,
\end{equation}
where $\epsilon = i \sigma_2$ and we adopt latin and greek letters for quark and lepton flavour indices, respectively. The weak-doublets quarks $q_i$ and leptons $\ell_\alpha$ are in the down-quark and charged-lepton  mass eigenstate bases. Note that Gauge invariance enforces antisymmetry of the $\phi^+$ couplings: $\lambda_{\alpha \beta} = - \lambda_{\beta \alpha}$. \par
It is worth mentioning that the LQ $S_1$ and $\phi^+$ share the same quantum numbers with those of a right-handed sbottom and stau. The couplings $\lambda^{1L}$ and $\lambda$ terms correspond then to the $\lambda'$ and $\lambda$ ones of the R-parity violating (RPV) superpotential~\cite{Barbier:2004ez}, respectively, while the couplings $\lambda^{1R}$ can potentially originate from non-holomorphic RPV terms~\cite{Csaki:2013jza}. The complete resolution to all the anomalies presented in this work may thus constitute a hint towards a RPV scenario with lighter 3rd generation superpartners~\cite{Trifinopoulos:2018rna,Trifinopoulos:2019lyo,Altmannshofer:2020axr}. \par 
Regarding the couplings employed in the analysis, we do not consider $\lambda^{1L(R)}$ couplings to the first generation quarks and leptons, as well as $\lambda_{s\mu}^{1L}$ and $\lambda_{t\tau}^{1R}$, which are not needed for the explanation of the anomalies. Moreover, we set $\lambda_{e\tau} \approx 0$ in order to satisfy the very strict constraints from the Lepton Flavour Violating (LFV) decay $\mu \to e \gamma$~\cite{Crivellin:2020klg}. We assume NP couplings to be real, for simplicity.

\begin{figure}[t]
\centering
  \begin{subfigure}{.2\textwidth}
    \includegraphics[width=1\linewidth]{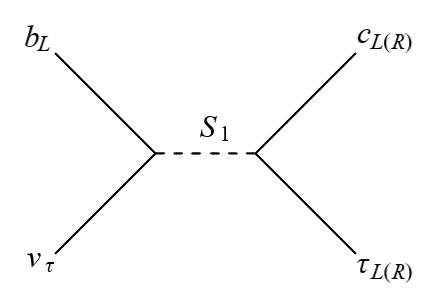}
    \caption{\label{fig:bctaunu}}
  \end{subfigure}%
  \begin{subfigure}{.2\textwidth}
    \includegraphics[width=1\linewidth]{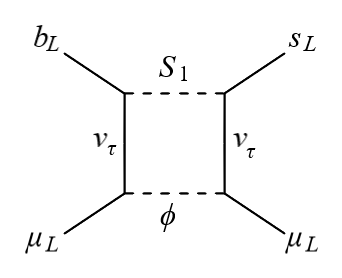}
    \caption{\label{fig:bsmumu}}
  \end{subfigure}
    \begin{subfigure}{.2\textwidth}
    \includegraphics[width=1\linewidth]{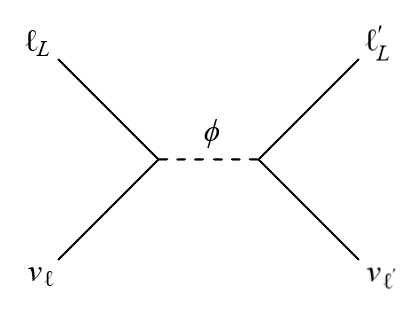}
    \caption{\label{fig:llpnunu_tree}}
  \end{subfigure}%
    \begin{subfigure}{.2\textwidth}
    \includegraphics[width=1\linewidth]{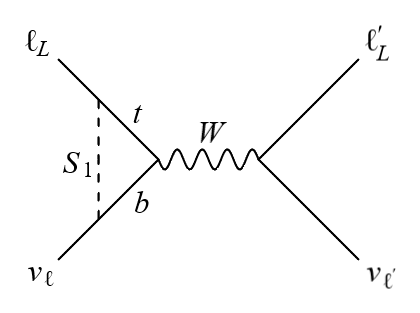}
    \caption{\label{fig:llpnunu_loop}}
  \end{subfigure}
      \begin{subfigure}{.2\textwidth}
    \includegraphics[width=1\linewidth]{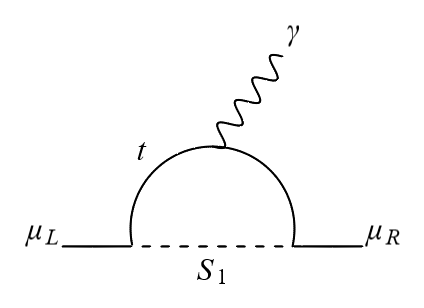}
    \caption{\label{fig:g-2}}
  \end{subfigure}
\caption{The diagrams that generate the dominant contributions to the flavour anomalies.}
\label{fig:anomalies_diag}
\end{figure}

\section{Observables}
\label{sec:observables}

In this section, we present the dominant contributions due to $S_1$ and $\phi^+$  to the anomalous observables. We obtain the $S_1$ contributions using the results of Ref.~\cite{Gherardi:2020qhc,Gherardi:2020det}, to which we refer for more details. In the numerical analysis the complete expressions are employed.

A tree-level $S_1$ exchange is invoked in order to explain $b \to c \tau \nu$ anomalies (see Fig. \ref{fig:bctaunu}). The approximate numerical expressions for the $R_{D^{(*)}}$ ratios relevant for the parameter region of interest are
\begin{eqnarray}
	&R_D \approx 0.299 - 0.235 \frac{\lambda^{1L}_{b\tau} \lambda^{1R}_{c\tau}}{m_1^2} \left( 1 + 0.05 \log m_1^2 \right) , \\
	&R_{D^*} \approx 0.258 - 0.088 \frac{\lambda^{1L}_{b\tau} \lambda^{1R}_{c\tau}}{m_1^2} \left( 1 + 0.02 \log m_1^2 \right) ,
\end{eqnarray}
where $m_1 \equiv M_1 / \TeV$. Note that quadratic terms and purely left-handed contributions are sub-leading in our setup. The logarithm becomes important for large masses and enhances the effect in $R_D$ compared to $R_{D^*}$.

The observables related to the \mbox{$b \to s \ell \ell$} anomalies receive contributions generated from the Wilson Coefficients (WCs) of the operators \mbox{$\mathcal{O}^{bs\mu\mu}_{LL (LR)} = (\bar{s} \gamma^\alpha P_L b)(\bar\mu \gamma_\alpha P_{L(R)} \mu)$}. They are given by (see also \cite{Bauer:2015knc})
\begin{eqnarray}
    \! C_{LL}&\approx& - \lambda^{1L}_{b\tau} \lambda^{1L\, *}_{s\tau} \left(\frac{|\lambda^{1L}_{b\mu}|^2 }{64 \pi^2 M_1^2} + \frac{|\lambda_{\mu\tau}|^2 \log M_\phi^2/M_1^2 }{64 \pi^2 (M_\phi^2 - M_1^2)}  \right) \label{eq:c_LL} \\
	\! C_{LR} &\approx& - \frac{|\lambda^{1R}_{c\mu}|^2 \lambda^{1L}_{b\tau} \lambda^{1L\, *}_{s\tau}}{64 \pi^2 M_1^2}~.
	\label{eq:c_LR}
\end{eqnarray}
The second term in Eq. \eqref{eq:c_LL} corresponds to the diagram in Fig. \ref{fig:bsmumu} and yields the leading contribution in this scenario.
Eventually, the results of the global fit are expressed in terms of the low-energy WCs in the standard notation \mbox{$\Delta C_{9,10}^\mu = (C_{LR} \pm C_{LL}) / (2 \mathcal{N}_{sb})$}, where \mbox{$\mathcal{N}_{sb} = \frac{G_F \alpha V_{tb} V_{ts}^*}{\sqrt{2} \pi}$}. \par

The leading $S_1$ contribution to the anomalous muon magnetic moment arises via a triangle diagram (see Fig. \ref{fig:g-2}) and is given by
\begin{equation}
	\Delta a_\mu \approx \frac{ m_\mu m_t  \lambda^{1L}_{b\mu} \lambda^{1R}_{t\mu}}{4 \pi^2 M_1^2} \left(\log M_1^2 / m_t^2 - \frac{7}{4} \right)~.
\end{equation}

\begin{table}[t]
\addtolength{\arraycolsep}{3pt}
\renewcommand{\arraystretch}{1.2}
\centering
\begin{tabular}{|c|c|}
\hline
Observable & Experimental value  \\
\hline\hline
$R_D$ &  $0.34 \pm 0.029$~\cite{Amhis:2019ckw}  \\
\hline
$R_{D^*}$ &  $0.295 \pm 0.013$~\cite{Amhis:2019ckw}  \\
\hline
\begin{tabular}{@{}l@{}}
                   $\Delta C_9^{\mu}$\\
                   $\Delta C_{10}^{\mu}$\\
                 \end{tabular} & 
\begin{tabular}{@{}l@{}}
                   $-0.675 \pm 0.16$~\cite{Altmannshofer:2021qrr} \\
                   $0.244 \pm 0.13$~\cite{Altmannshofer:2021qrr} \\
                 \end{tabular}  \\ 
\hline
$\Delta a_{\mu}$ &  $(2.51 \pm 0.59)\times 10^{-9}$~\cite{Abi:2021gix,Aoyama:2020ynm} \\
\hline
$\delta(\mu \to e \nu\nu)$ & $(6.5 \pm 1.5)\times 10^{-4}$~\cite{Crivellin:2020klg}  \\
\hline
$R_D^{\mu/e}$ &  $0.978 \pm 0.035$~\cite{Aubert:2008yv,Glattauer:2015teq}  \\
\hline
$\Br(B_c \to \tau \nu)$ & $< 0.1$ \cite{Akeroyd:2017mhr}   \\
\hline
$R_{K^{(*)}}^{\nu}$ & $<2.7$~\cite{Grygier:2017tzo}  \\
\hline
$C_{B_s}^1$ & $<2.01 \times 10^{-5}~\TeV^{-2}$~\cite{Bona:2007vi}  \\
\hline
$|\text{Re}(C_{D}^1)|$ &  $<3.57 \times 10^{-7}~\TeV^{-2}$~\cite{Bona:2007vi}  \\
\hline
$|\text{Im}(C_{D}^1)|$ &  $<2.23 \times 10^{-8}~ \TeV^{-2}$~\cite{Bona:2007vi}  \\
\hline
$\frac{g_{\tau}}{g_{e}}$ & $1.0058 \pm 0.0030$~\cite{Amhis:2019ckw}  \\
\hline
$\frac{g_{\tau}}{g_{\mu}}$ & $1.0022 \pm 0.0030$~\cite{Amhis:2019ckw}  \\
\hline
$\frac{g_{\mu}}{g_{e}}$ & $1.0036 \pm 0.0028$~\cite{Amhis:2019ckw}  \\
\hline
$\delta g^Z_{\tau_L}$ & $(-0.11 \pm 0.61) \times 10^{-3}$~\cite{ALEPH:2005ab}  \\
\hline
$\delta g^Z_{\tau_R}$ & $(0.66 \pm 0.65) \times 10^{-3}$~\cite{ALEPH:2005ab}  \\
\hline
$\delta g^Z_{\mu L}$ & $(0.3 \pm 1.1) \times 10^{-3}$~\cite{ALEPH:2005ab}  \\
\hline
$\delta g^Z_{\mu R}$ & $(0.2 \pm 1.3) \times 10^{-3}$~\cite{ALEPH:2005ab}  \\
\hline
$\mathcal B(\tau \to \mu \gamma)$ & $<4.4 \times 10^{-8}$~\cite{Patrignani:2016xqp}  \\
\hline
$\mathcal B(\tau \to 3 \mu)$ & $<2.1 \times 10^{-8}$~\cite{Patrignani:2016xqp} \\ \hline
\end{tabular}
\caption{Experimental values for the observables used in the numerical analysis. In case of $R_{D^{(*)}}$, $\Delta C_{9,10}^\mu$, and $\tau$ LFU the relevant correlations are taken into account.}
\label{tbl:obs}
\end{table}

The presence of $\phi^+$ at tree-level (see Fig. \ref{fig:llpnunu_tree}) and $S_1$ at one loop (see Fig. \ref{fig:llpnunu_loop}) implies the following NP effects in the charged-current muon decay:
\begin{equation}\label{eq:muenunu}
    \delta(\mu \to e \nu\nu) \approx \frac{v^2 |\lambda_{12}|^2}{4 M_\phi^2} + \frac{3 m_t^2 |\lambda^{1L}_{b\mu}|^2}{32 \pi^2 M_1^2} \left(  \frac{1}{2} - \log \frac{M_1^2}{m_t^2}\right)~,
\end{equation}
where $\delta(\ell \to \ell^\prime \nu\nu) \equiv \mathcal{A}(\ell \to \ell^\prime \nu\nu)_{\rm NP} / \mathcal{A}(\ell \to \ell^\prime \nu\nu)_{\rm SM}$. \par
As investigated in Ref.~\cite{Belfatto:2019swo,Crivellin:2020klg}, one can alleviate the tension between the value of $V_{us}$ computed from Kaon decays,\footnote{This is an average of the value extracted from $K \to \pi \ell \nu$ decays $V_{us}^{K \ell3}=0.22326(58)$ and the ratio $\mathcal B(K \to \mu \nu)/\mathcal B(\pi \to \mu\nu)$, $V_{us}^{K\mu 2}=0.22534(42)$~\cite{Aoki:2019cca}. Note that the discrepancy between $V_{us}^{K \ell3}$ and $V_{us}^{K \mu 2}$ cannot be explained by LFU violation.}
$V_{us}^{\rm CKM}=0.2243(5)$ and the one computed via CKM unitarity from $V_{ud}^{\beta}$ as extracted from nuclear beta-decays \cite{Seng:2020wjq}, i.e. $V_{us}^{\beta}=0.2280(6)$, by introducing a constructive interference in $\mu \to e \nu \nu$. In particular, one obtains:
\begin{align}
    V_{us}^{\beta} &\equiv \sqrt{1-(V_{ud}^{\beta})^2-|V_{ub}|^2} \notag \\
    &\simeq V_{us}^{\rm CKM}\left[ 1-\left( \frac{V_{ud}^{\rm CKM}}{V_{us}^{\rm CKM}} \right)^2\delta(\mu \to e \nu\nu) \right]~,
\end{align}
where $V_{ud}^{\rm CKM}=0.97420(21)$ and $|V_{ub}|^2\approx 10^{-5}$~\cite{Tanabashi:2018oca} is negligible. Eventually, a global fit including the standard EW observables yields the value of $\delta(\mu \to e \nu\nu)$ indicated at Table~ \ref{tbl:obs} (see Refs. \cite{Belfatto:2019swo,Crivellin:2021njn} for details). \par

\begin{figure*}[t]
\centering
\includegraphics[height=5.5cm]{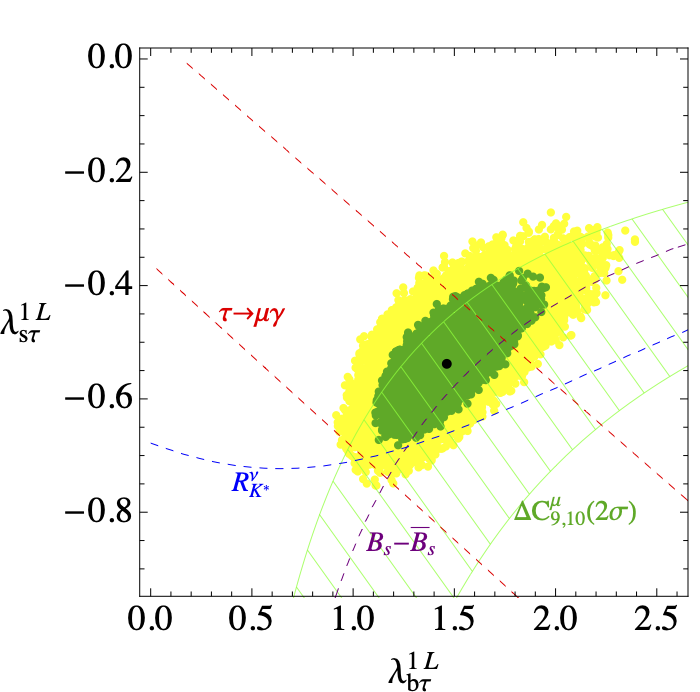} 
\includegraphics[height=5.5cm]{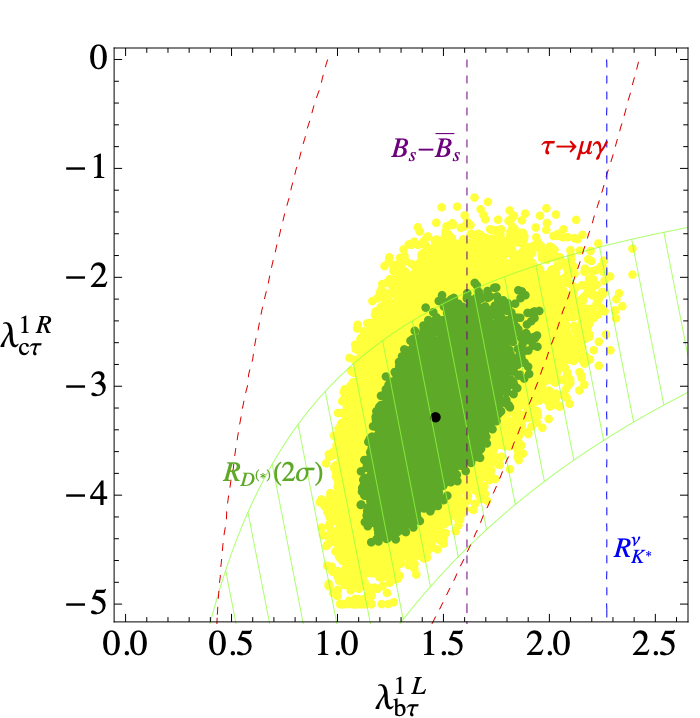}
\includegraphics[height=5.5cm]{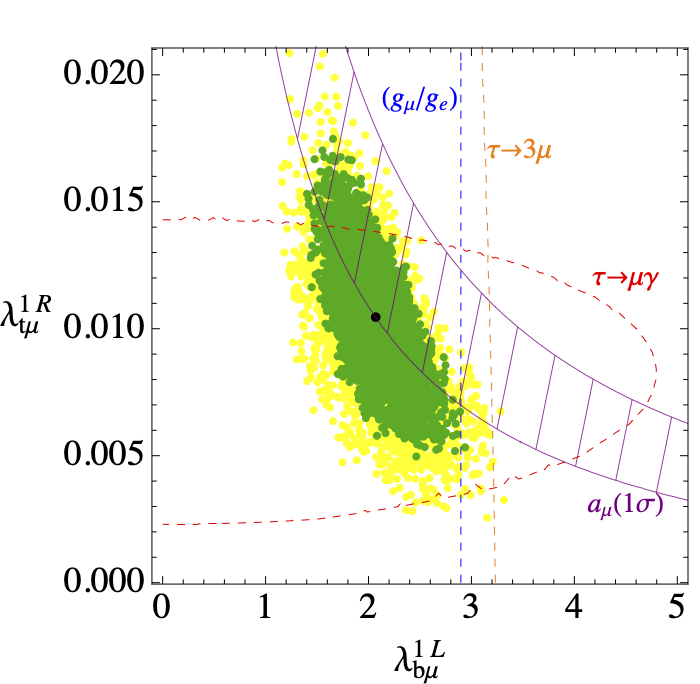} \\[0.5cm]
\includegraphics[height=5cm]{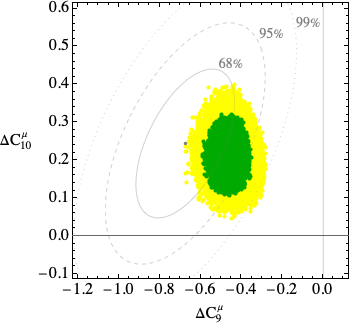} 
\includegraphics[height=5.3cm]{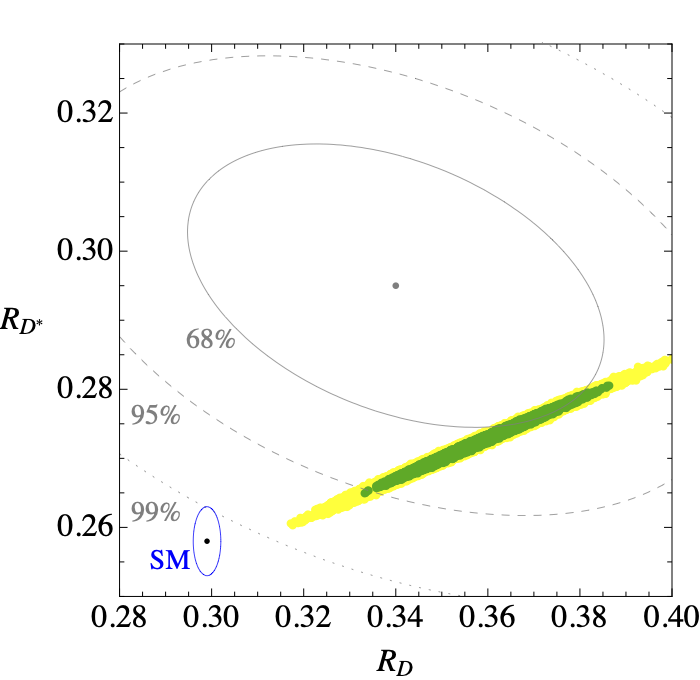} 
\includegraphics[height=5cm]{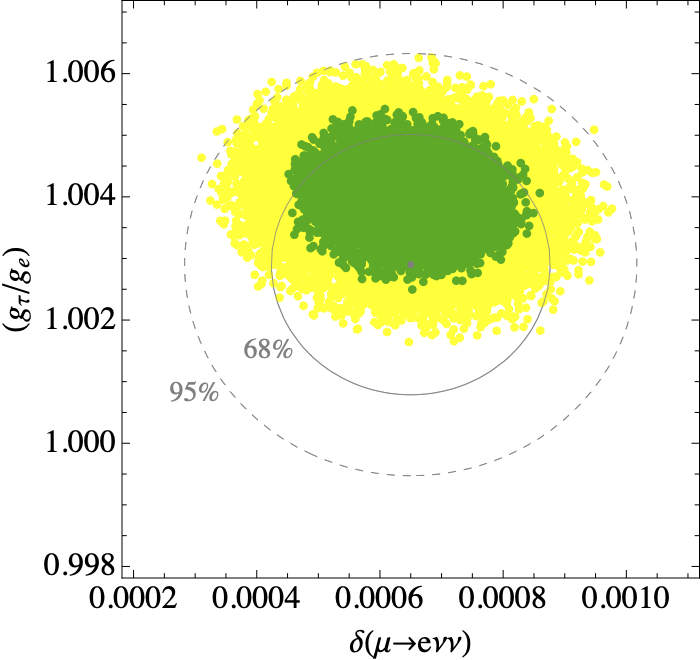} 
\caption{\label{fig:scan} Results of the parameter scan of the model's parameters, with $M_1 = M_\phi = 5.5 \TeV$. The green (yellow) points are within $1\sigma$ ($2\sigma$) of the best-fit point, shown in black. The upper row shows the preferred region for some of the couplings and the single-observable 95\%CL constraints. The bottom row shows how this preferred region maps in the plane of pairs of observables of interest.}
\end{figure*}

\section{Phenomenology} 
\label{sec:Phenomenology}

\emph{Global analysis --} With all the observables listed in Table~\ref{tbl:obs} and the expressions for the $S_1$ and $\phi^+$ contributions given in Sec. \ref{sec:observables} and in the Appendix, we build a global likelihood $\chi^2 = - 2 \log \mathcal{L}$. We find the best-fit point by minimizing the $\chi^2$ function and compare it to the value obtained in the SM.

This analysis prefers large values for the scalar masses $M_1$ and $M_\phi$. This can be understood by the fact that the contributions to $b\to s \mu\mu$ scale as $\lambda^4 / M^2$, while most constraints scale as $\lambda^2 / M^2$, except for $B_s$-mixing, which scale with $\lambda^4 / M^2$. Larger masses, and couplings, allow thus to better fit the neutral-current B-anomalies and remain compatibile with the other constraints.
On the other hand, in order to avoid too large couplings, that would put the perturbativity of the model into question, the masses cannot be too large.

Fixing $M_1 = M_\phi = 5.5 \TeV$ (we chose equal masses only for simplicity), we find the following best-fit point:
\begin{equation}
    \begin{array}{lll}
        \lambda_{e\mu} = 1.35~, &
        \lambda_{\mu\tau} = 3.17~, \\
        \lambda^{1L}_{b\tau} = 1.46~, &
        \lambda^{1L}_{s\tau} = -0.54~, &
        \lambda^{1L}_{b\mu} = 2.07~, \\
        \lambda^{1R}_{c\tau} = -3.28~, &
        \lambda^{1R}_{t\mu} = 0.01~, & 
        \lambda^{1R}_{c\mu} = 2.35~,
    \end{array}
\end{equation}
for which $\chi^2_{\rm SM} - \chi^2_{\rm best-fit} = 82$, which constitutes a major improvement from the SM.
The coupling $\lambda^{1R}_{c\mu}$ is required to cancel an otherwise excessive contribution to $\tau \to \mu \gamma$. The required cancellation in the amplitude (see Appendix) is approximately of one part in three.

To study the preferred region in parameter space we perform a numerical scan via a Markov-Chain Monte Carlo algorithm that we use to select points with $\Delta \chi^2 = \chi^2 - \chi^2_{\rm best-fit}$ corresponding to 68\% and 95\% confidence level (CL) regions.
The results of this scan are shown in Fig.~\ref{fig:scan}. In the top row we show the preferred regions for pairs of couplings as well as the relevant single-observable constraints in each plane.\footnote{For brevity we don't show analogous plots for $\lambda^{1R}_{c\mu}$, which has values in the interval $[1.5, 3]$, and the $\phi^+$ couplings, which take values $\lambda_{e\mu} \in [1.1, 1.6]$ and $\lambda_{\mu \tau} \in [2.7, 3.6]$.}
In the bottom row we show how these preferred regions map into pairs of the observables showing a discrepancy with the SM (the effect in $\Delta a_\mu$  can be seen in the top-right plot comparing with the purple-meshed region representing the experimentally preferred value at $1\sigma$).

We observe that the model is able to address at the $1\sigma$ level all the four deviations from the SM presented in the Introduction. As a byproduct, a small tension present in LFU tests in $\tau$ decays, $(g_\tau/g_e)$, is also addressed in this framework.

\begin{figure}[t]
\centering
\includegraphics[height=6cm]{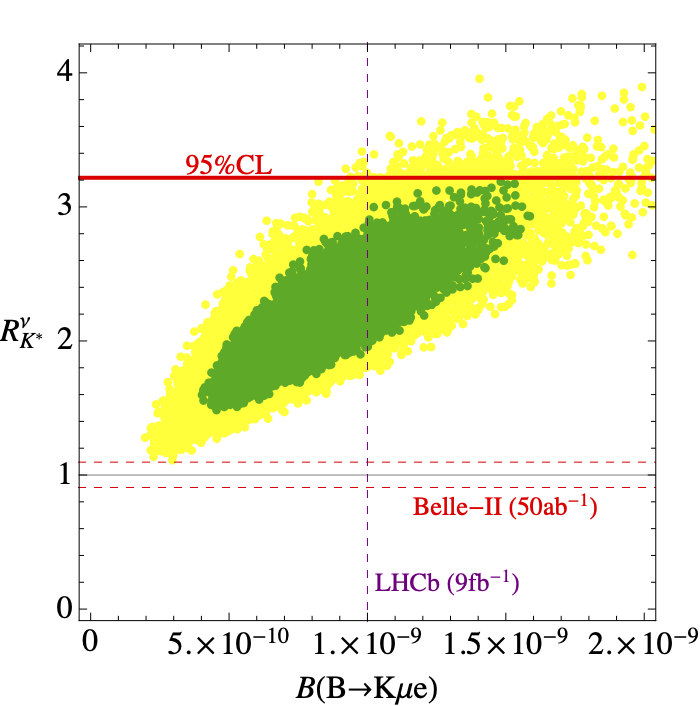} 
\caption{\label{fig:prospects} Here the preferred values of $\Br(B\to K \mu e)$ and $R_{K^*}^\nu$ are shown, together with the present 95\% CL limit (red line) and the future prospects expected by LHCb~\cite{Bediaga:2018lhg} and Belle II~\cite{Kou:2018nap}.
}
\end{figure}

\emph{Future prospects --} 
Both $B_s$-mixing and $B\to K^{(*)}\nu\nu$ are sensitive to the $S_1$ couplings contributing to $R_{K^{(*)}}$ and $R_{D^{(*)}}$, and the preferred region by the model is close to the present exclusion limits, as shown in Fig.~\ref{fig:prospects} for $R^\nu_{K^{*}}$. A deviation from the SM could thus reveal itself in future updates of this observable by the Belle-II experiment \cite{Kou:2018nap}.

Via a one-loop box diagram with both $S_1$ and $\phi^+$, similar to Fig.\ref{fig:anomalies_diag}(b), a contribution to the LFV process \mbox{$b\to s \mu e$} is induced. The preferred values in our model for \mbox{$\Br(B \to K \mu e)$} are shown in Fig.~\ref{fig:prospects}, while \mbox{$\Br(B \to K^* \mu e) \approx 2.1 \, \Br(B \to K \mu e)$} and \mbox{$\Br(B_s \to \mu e) \sim \mathcal{O}(10^{-12})$}.
On the other hand, due to the specific structure of the couplings, in this model we do not predict sizeable effects in $b\to s \tau \tau$ and $b \to s \tau \mu$ processes.

As shown in Fig.~\ref{fig:scan} (bottom-right) we also expect per-mille effects in LFU tests in $\tau$ decays, which is in the range of future sensitivity by Belle-II  \cite{Kou:2018nap}.
The model predicts also effects in LFV $\tau$ decays. The $S_1$ LQ generates \mbox{$\tau \to \mu \gamma$}, \mbox{$\tau \to 3 \mu$}, and \mbox{$\tau \to \mu e e$} with rates close to the present bounds (of the order of $\sim 10^{-8}$). The scalar $\phi^+$, instead, mediates \mbox{$\Br(\tau \to e \mu\mu) \sim 10^{-9}$}, \mbox{$\Br(\tau \to 3 e) \sim 10^{-10}$},  and \mbox{$\Br(\tau \to e \gamma) \sim 10^{-11}$} \cite{Crivellin:2020klg}. Also for these channels Belle-II and LHCb are expected to improve substantially on the present constraints by at least one order of magnitude \cite{Bediaga:2018lhg,Kou:2018nap}.

Finally, while the large masses preferred by the fit are beyond the reach of direct searches at LHC, effects in high-energy tails of Drell-Yan due to $S_1$ are possible. At FCC-hh the leptoquark could be produced on-shell and a muon collider would be the ideal machine to study also the scalar $\phi^+$.

\section{Conclusions}

In this letter we propose a New Physics model addressing  the most significant deviations from the SM observed in flavour physics, while being at the same time consistent with all phenomenological constraints.
The model is the first one that establishes a connection between all four classes of flavour anomalies under the same LFU violating interpretation. Furthermore, since it comprises of only two weak-singlet scalars: the leptoquark $S_1$ and the colorless $\phi^+$, it is also the most minimal solution to be proposed in the literature for a combined resolution of them.

In the foreseeable future, the LHCb and Belle-II experiments will clarify the nature of the present anomalies in $B$-decays, while the Fermilab $(g-2)_\mu$ experiment has already been collecting a large amount of additional data that will allow to further reduce the experimental uncertainty.
In order to settle the CKM unitarity puzzle, experimental developments are expected in the existing precision observables used for the determination of the Cabibbo angle~\cite{Crivellin:2020lzu} as well as further observables such as hadronic $\tau$ decays~\cite{Lusiani:2018zvr}, the pion $\beta$ decay~\cite{Czarnecki:2019iwz} and the neutron lifetime~\cite{Czarnecki:2019mwq} that can provide complementary tests in the future.

If any one of these signals will be further confirmed by future data it would imply a revolution in our understanding of fundamental interactions. However, it is only by the combination of several deviations in different observables that we might be able to pinpoint the precise nature of the underlying New Physics.

\subsection*{Acknowledgements}

The authors acknowledge support by MIUR grant PRIN 2017L5W2PT.
DM is also partially supported by the INFN grant SESAMO and by the European Research Council (ERC) under the European Union’s Horizon 2020 research and innovation programme, grant agreement 833280 (FLAY).

\appendix
\setcounter{section}{-1}
\renewcommand{\theequation}{A.\arabic{equation}}
\section{Details on the constraints}
\label{app:constraints}

Approximate expressions for the observables listed in Table~\ref{tbl:obs} are provided here. Unless stated otherwise, they have been taken from Ref.~\cite{Gherardi:2020qhc}, to which we refer for more details. In those cases where analytic formulas are not available or too complicated, we report approximate numerical expressions.

\emph{-- Meson mixing.}
The contribution to $B_s$ and $D^0$ meson mixing arises via the operators $\mathcal{O}_{B_s}^1 = (\bar s \gamma_\mu P_L b)^2$ and $\mathcal{O}_{D}^1 = (\bar u \gamma_\mu P_L c)^2$, with coefficients
\begin{equation}
	C^1_{B_s} = \frac{(\lambda^{1L\,*}_{b\tau} \lambda^{1L}_{s\tau} )^2}{128\pi^2 M_1^2}~, \quad
	C^1_{D} = \frac{(V_{ci} \lambda^{1L\,*}_{i\alpha} \lambda^{1L}_{j\alpha} V_{uj}^* )^2}{128\pi^2 M_1^2}~.
\end{equation}

\emph{-- $B \to K^{(*)} \nu\nu$.}
The $S_1$ couplings to left-handed fermions also contribute to the decays $B \to K^{(*)} \nu\nu$, the leading dependence being
\begin{equation}
    R^\nu_{K^{(*)}} \approx 1 + 34 \frac{ \lambda^{1L}_{s\tau}\lambda^{1L}_{b\tau} }{m_1^2} + 856 \frac{ (\lambda^{1L}_{s\tau})^2((\lambda^{1L}_{b\mu})^2+(\lambda^{1L}_{b\tau})^2) }{m_1^4},
\end{equation}
where $m_1 = M_1 / \TeV$ and $R^\nu_{K^{(*)}}$ is defined as the ratio of the branching ratio to the corresponding SM prediction. 

\emph{-- $B_c \to \tau \nu$.}
The branching ratio of $B_c \to \tau \nu$ is a sensitive probe to scalar operators contributing to \mbox{$b \to c \tau \nu$}, as the one induced by $S_1$. It is given approximately by
\begin{equation}
    \Br(B_c \to \tau \nu) \approx 0.02 + 0.12 \frac{\lambda^{1L}_{b\tau} \lambda^{1R}_{c\tau}}{m_1^2} \left(1 + 0.04 \log m_1^2 \right)~.
\end{equation}

\emph{-- LFU in $B \to D \ell \nu$.}
The large value of the $\lambda^{1L}_{b\mu}$ and $\lambda^{1R}_{c\mu}$ couplings required to fit $\Delta a_\mu$ and to cancel an excessive contribution to $\tau \to \mu \gamma$, respectively, could induce a too large deviation in the LFU ratio \mbox{$R_D^{\mu/e} \equiv \text{Br}(B \to D \mu \nu)/\text{Br}(B \to D e \nu)$}. The leading dependence on the model parameters is given by
\begin{eqnarray}
    R_D^{\mu/e} &\approx& 1 + 0.03 \frac{(\lambda^{1L}_{b\mu})^2}{m_1^2} - 0.047 \frac{\lambda^{1L}_{b\mu} \lambda^{1R}_{c\mu}}{m_1^2} + \nonumber \\
    && + 0.5 \frac{(\lambda^{1L}_{b\mu})^2 (\lambda^{1R}_{c\mu})^2}{m_1^4}~.
\end{eqnarray}

\emph{-- Charged-current lepton decays.} In addition to Eq.~\eqref{eq:muenunu}, one obtains the following modifications to charged-current leptonic decays~\cite{Gherardi:2020qhc,Crivellin:2020klg}:
\begin{eqnarray}
    \delta(\tau \to \mu \nu\nu) &\approx& \frac{v^2 |\lambda_{\mu\tau}|^2}{4 M_\phi^2} + \frac{3 v^2 |\lambda^{1L}_{b\mu}|^2 |\lambda^{1L}_{b\tau}|^2}{128 \pi^2 M_1^2} + \nonumber\\
        && + \frac{3 m_t^2 (|\lambda^{1L}_{b\mu}|^2 + |\lambda^{1L}_{b\tau}|^2)}{32 \pi^2 M_1^2} \left(  \frac{1}{2} - \log \frac{M_1^2}{m_t^2}\right)~, \nonumber\\
    \delta(\tau \to e \nu\nu) &\approx& \frac{3 m_t^2 |\lambda^{1L}_{b\tau}|^2}{32 \pi^2 M_1^2} \left(  \frac{1}{2} - \log \frac{M_1^2}{m_t^2}\right)~.
\end{eqnarray}

The LFU ratios in $\tau$ decays~\cite{Amhis:2019ckw} are then given by
\begin{eqnarray}
    \frac{g_\tau}{g_e} &=& \left| \frac{1 + \delta(\tau \to \mu \nu\nu)}{1 + \delta(\mu \to e \nu\nu)} \right|~, \quad
    \frac{g_\tau}{g_\mu} = \left| \frac{1 + \delta(\tau \to e \nu\nu)}{1 + \delta(\mu \to e \nu\nu)} \right|~, \nonumber\\
    \frac{g_\mu}{g_e} &=& \left| \frac{1 + \delta(\tau \to \mu \nu\nu)}{1 + \delta(\tau \to e \nu\nu)} \right|~.
\end{eqnarray}

\emph{-- LFV $\tau$ decays.}
In our framework, box and penguin diagrams involving $S_1$ in the loop generate NP contributions to the operators \mbox{$\mathcal{O}_{LL(R)}^{3\mu}= (\bar{\tau}_L \gamma^{\mu} \mu_L)(\bar{\mu}_{L(R)} \gamma_{\mu} \mu_{L(R)})$}. 

The most important terms to the respective WCs are
\begin{align}
C_{LL}^{3\mu} &\approx \left( \frac{1}{2} - c_W^2 \right) 3 y_t^2 \frac{\lambda_{b \mu}^{1L*} \lambda_{b \tau}^{1L}}{2 M_1^2} \left( 1+\log\frac{m_t^2}{M_1^2} \right)~, \\
C_{LR}^{3\mu} &\approx 2\left( 1 - c_W^2 \right) 3 y_t^2 \frac{\lambda_{b \mu}^{1L*} \lambda_{b \tau}^{1L}}{2 M_1^2} \left( 1+\log\frac{m_t^2}{M_1^2} \right)~.
\end{align}
We also compute the form factors that parametrize the radiative $\tau$ decays
\begin{align}
T^{R}_{\mu\tau} \approx& - \frac{e m_{c}}{8\pi^2} \frac{ V_{cb} \lambda_{b\mu}^{1L*}\lambda_{c\tau}^{1R}}{M_{1}^{2}} \left( \log \frac{m_c^2}{M_1^2} + \frac{7}{4} \right) + \frac{e m_{\tau} }{64\pi^2}  \frac{\lambda^{1L \dagger}_{b\mu} \lambda^{1L}_{b\tau}}{M_1^2}~, \notag \\
T^{L}_{\mu\tau} \approx& - \frac{e m_{t}}{8\pi^2} \frac{ \lambda_{b\tau}^{1L*}\lambda_{t\mu}^{1R}+V_{ts}\lambda_{s\tau}^{1L*}\lambda_{t\mu}^{1R}}{M_{1}^{2}} \left( \log \frac{m_t^2}{M_1^2} + \frac{7}{4} \right) \notag \\
&- \frac{e m_{c}}{8\pi^2} \frac{ V_{cb} \lambda_{b\tau}^{1L*}\lambda_{c\mu}^{1R}}{M_{1}^{2}} \left( \log \frac{m_c^2}{M_1^2} + \frac{7}{4} \right) +\frac{e m_{\tau} }{64\pi^2}  \frac{\lambda^{1R \dagger}_{c\tau} \lambda^{1R}_{c\mu}}{M_1^2}~. 
\end{align}
The branching ratios~\cite{Crivellin:2013hpa},
\begin{align}
\label{eq:Btauto3mu}
&\mathcal B(\tau \to 3 \mu) \approx \frac{m_{\tau}^5}{3(16\pi)^2 \Gamma_{\tau}} \left( \left|C_{LL}^{3\mu}\right|^2+\frac{1}{2}\left|C_{LR}^{3\mu}\right|^2 \right) \\
&\mathcal B(\tau \to \mu \gamma) = \frac{m_{\tau}^3}{16 \pi \Gamma_{\tau}} (\left|T^{R}_{\mu\tau}\right|^2 + \left|T^{L}_{\mu\tau}\right|^2)~,
\end{align}
must then comply with the respective experimental upper bounds.

\emph{-- $Z$ boson couplings.}
Triangle diagrams with $S_1$ in the loop modify the $Z$-boson couplings as:
\begin{eqnarray}
    10^3 \delta g^Z_{e_\alpha L} &\approx& 0.59 \frac{(\lambda^{1L}_{b\alpha})^2}{m_1^2} \left(1 + 0.39 \log m_1^2 \right)~, \nonumber\\
    10^3 \delta g^Z_{e_\alpha R} &\approx& -0.67 \frac{(\lambda^{1R}_{t\alpha})^2}{m_1^2} \left(1 + 0.36 \log m_1^2\right) + \nonumber \\
    && + 0.06 \frac{(\lambda^{1R}_{c\alpha})^2}{m_1^2} \left(1 + 0.14 \log m_1^2\right)~.
\end{eqnarray}

{\small
\bibliographystyle{apsrev4-2}
}

\end{document}